\documentstyle[12pt]{article}
\newcommand\be{\begin{equation}}
\newcommand\ee{\end{equation}}
\newcommand\bea{\begin{eqnarray}}
\newcommand\eea{\end{eqnarray}}
\newcommand\ket[1]{|#1\rangle}

\newcommand{\fatalpha}{{\bf \alpha \kern -0.44em \alpha}}
\newcommand{\fatsigma}{{\bf \sigma \kern -0.54em \sigma}}
\newcommand{\tpchi}{{\bf \chi \kern -0.35em \chi}}
\newcommand{\llambda}{{\bf \lambda \kern -0.45em \lambda}}

\bibliography{plain}
\title{\bf Entanglement entropy in the Hypercube networks}\vspace{20mm}
\author{ M. A. Jafarizadeh$^{a}$
 \thanks{E-mail:jafarizadeh@tabrizu.ac.ir},
 F. Eghbalifam$^{a}$
 \thanks{E-mail:F.Egbali@tabrizu.ac.ir},
 S. Nami$^{a}$
 \thanks{E-mail:S.Nami@tabrizu.ac.ir}
\\ $^a${\small Department of Theoretical Physics and Astrophysics,
University of Tabriz, Tabriz 51664, Iran.}} \pagebreak

\vspace{20mm}
\begin{document}
\maketitle \vspace{15mm}
\begin{abstract}
We investigate the hypercube networks that their nodes are
considered as quantum harmonic oscillators. The entanglement of
the ground state can be used to quantify the amount of information
each part of a network shares with the rest of the system via
quantum fluctuations. Therefore, the Schmidt numbers and
entanglement entropy between two special parts of Hypercube
network, can be calculated. To this aim, first we use the
stratification method to rewrite the adjacency matrix of the
network in the stratification basis which is the matrix
representation of the angular momentum. Then the entanglement
entropy and Schmidt number for special partitions are calculated
analytically by using the generalized Schur complement method.
Also, we calculate the entanglement entropy between two arbitrary equal subsets (two equal subsets have the same number of vertices) in $H(3,2)$  and $H(4,2)$ numerically, and we give
the minimum and maximum values of entanglement entropy in these two Hypercube network. Then we can conjecture the minimum and maximum values of entanglement entropy for equal subsets in $H(d,2)$.
\end{abstract}
\section{Introduction}
Entanglement plays a crucial role in quantum information
processing, including quantum communication [1,2] and quantum
computation [3-5]. It is one of the remarkable features that
distinguishes quantum mechanics from classical mechanics.\\ For
decades, entanglement has been the focus of much work in the
foundations of quantum mechanics, being associated particularly
with quantum nonseparability and the violation of Bells
inequalities [6]. Since entanglement has become regarded as such
an important resource, there is a need for a means of quantifying
it. For the case of bipartite entanglement, a recent exhaustive
review was written by the Horodecki family [7] and entanglement
measures have been reviewed in detail by Virmani and Plenio [8].
One of the operational entanglement criteria is the Schmidt
decomposition [9-11]. The Schmidt decomposition is a very good
tool to study entanglement of bipartite pure states. The Schmidt
number provides an important variable to classify entanglement.
The entanglement of a partly entangled pure state can be naturally
parametrized by its entropy of entanglement, defined as the von
Neumann entropy, or equivalently as the Shannon entropy of the
squares of the Schmidt coefficients [9,11]. The situation
simplifies if only so called \textit{Gaussian states} of the
harmonic oscillator modes are considered [12-17]. The importance
of gaussian states is two-fold; firstly, its structural
mathematical description makes them much more amenable than any
other continuous variable system (Continuous variable systems are
those described by canonical conjugated coordinates $x$  and $p$
endowed with infinite dimensional Hilbert spaces). Secondly, its
production, manipulation and detection with current optical
technology can be done with a very high degree of accuracy and
control. In [18] the authors quantified the amount of information
that a single element of a quantum network shares with the rest of
the system. They considered a network of quantum harmonic
oscillators and analyzed its ground state to compute the entropy
of entanglement that vacuum fluctuations creates between single
nodes and the rest of the network by using the Von Neumann
entropy. In [20], jafarizadeh et al, quantify the entanglement
entropy between two parts of the network. To this aim, they compute
the vacuum state of bosonic modes harmonically coupled through the
specific adjacency matrix of a given network. \\In this paper, first we rewrite the adjacency matrix of Hypercube network in the stratification basis, and we show that it is the matrix representation of angular momentum. Then we calculate the Schmidt numbers and entanglement entropy
between two subsets of Hypercube network by using the generalized
Schur complement method in three kinds of partitioning. Then we calculate the entanglement entropy between two equal subsets in $H(3,2)$ and $H(4,2)$ numerically, and we give the minimum and maximum values of entanglement entropy.
\\In  section II, we give some
preliminaries such as definitions related to association schemes,
corresponding stratification and The Terwilliger algebra.
\\In section III, the generalized Schur complement method is used for calculating the Schmidt numbers and entanglement entropy.
In this method, we will apply the generalized Schur complement
method to the potential matrix in the stratification basis several
times to calculate the entanglement entropy between two equal
parts in the graph. Therefore, by using this method, we can
calculate bipartite entanglement entropy between two equal parts of
Hypercube network for different partitions.
\\In section IV, we separate the vertices of $H(3,2)$ and $H(4,2)$ network into to equal subsets, then calculate the entanglement entropy between all kinds of
bisections. Then, we conjecture the maximum and minimum values of entanglement entropy
in H(d,2).

\section{Preliminaries} In this section we give some
preliminaries such as definitions related to association schemes,
corresponding stratification.

\subsection{The model and Hamiltonian}
We consider nodes as identical quantum oscillators, interacting
as dictated by the network topology encoded in the Laplacian $L$.
The Laplacian of a network is defined from the Adjacency matrix
as $L_{ij} = k_i\delta_{ij}- A_{ij}$ , where $k_i =\sum_j A_{ij}$
is the connectivity of node $i$, i.e., the number of nodes
connected to $i$. The Hamiltonian of the quantum network thus
reads:
\begin{equation}
H=\frac{1}{2}(P^T P+ X^T(I+2gL)X)
\end{equation}
here $I$ is the $N \times N$ identity matrix, $g$ is the coupling
strength between connected oscillators while $p^T=(p_1,p_2,...,
p_N)$ and $x^T=(x_1,x_2, ..., x_N)$ are the operators
corresponding to the momenta and positions of nodes respectively,
satisfying the usual commutation relations: $[x, p^T] = i\hbar I$
(we set $\hbar = 1$ in the following) and the matrix $V=I+2gL$ is
the potential matrix. Then the ground state of this Hamiltonian
is:
\begin{equation}
\psi(X)=\frac{(det(I+2gL))^{1/4}}{\pi^{N/4}}exp(-\frac{1}{2}(X^T(I+2gL)X))
\end{equation}
Where the $A_g=\frac{(det(I+2gL))^{1/4}}{\pi^{N/4}}$ is the
normalization factor for wave function. The elements of the
potential matrix in terms of entries of adjacency matrix is
$$V_{ij}=(1+2g\kappa_i)\delta_{ij}-2gA_{ij}$$

\subsection{Schmidt decomposition and entanglement entropy}

Any bipartite pure state $|\psi\rangle_{AB} \in
\textsl{H}=\textsl{H}_A \otimes\textsl{H}_B$ can be decomposed, by
choosing an appropriate basis, as
\begin{equation}
|\psi\rangle_{AB}=\sum_{i=1}^m \alpha_i|a_i\rangle\otimes|b_i\rangle
\end{equation}
where $1 \leq m \leq min\{dim(\textsl{H}_A); dim(\textsl{H}_B)\}$,
and $\alpha_i
> 0$ with $\sum_{i=1}^m \alpha_i^2 = 1$. Here $|a_i\rangle$ ($|b_i\rangle$) form a part of an
orthonormal basis in $\textsl{H}_A$ ($\textsl{H}_B$). The positive
numbers $\alpha_i$ are called the Schmidt coefficients of
$|\psi\rangle_{AB}$ and the number $m$ is called the Schmidt rank
of $|\psi\rangle_{AB}$.

Entropy of entanglement is defined as
the von Neumann entropy of either $\rho_A$ or $\rho_B$:
\begin{equation}
E=-Tr\rho_A log_2\rho_A= Tr\rho_B log_2\rho_B=-\sum_i\alpha_i^2
log_2 \alpha_i^2
\end{equation}

\subsection{Association scheme}
First we recall the definition of association schemes. The reader
is referred to Ref.[20], for further information on association
schemes.

\textbf{Definition 2.1} (Symmetric association schemes). Let $V$
be a set of vertices, and let $R_i(i = 0, 1, ..., d)$ be nonempty
relations on $V$ (i.e., subset of $V\times V$ ). Let the
following conditions (1), (2), (3) and (4) be satisfied. Then,
the relations $\{R_i\}_{0\leq i\leq d}$ on $V\times V$ satisfying
the following conditions

(1) $\{R_i\}_{0\leq i\leq d}$ is a partition of $V\times V$

(2) $R_0 = {(\alpha,\alpha) : \alpha \in V }$

(3) $R_i = R^t_i$ for $0\leq i \leq d$, where
$R^t_i={(\beta,\alpha) : (\alpha,\beta)\in R_i}$

(4) For $(\alpha,\beta)\in R_k$, the number $p^k_{ij}=|{\gamma
\in V : (\alpha,\gamma)\in R_i \quad and \quad (\gamma,\beta)\in
R_j}|$ does not depend on $(\alpha,\beta)$ but only on $i,j$ and
$k$, define a symmetric association scheme of class $d$ on $V$
which is denoted by $Y = (V, \{R_i\}_{0\leq i\leq d})$.
Furthermore, if we have $p^k_{ij} = p^k_{ji}$ for all $i, j, k =
0, 2, ..., d$, then $Y$ is called commutative.

The number $v$ of the vertices, $|V|$, is called the order of the
association scheme and $R_i$ is called $i$-th relation.

The intersection number $p^k_{ij}$ can be interpreted as the
number of vertices which have relation $i$ and $j$ with vertices
$\alpha$ and $\beta$, respectively provided that $(\alpha,\beta)
\in R_k$, and it is the same for all elements of relation $R_k$.
For all integers $i (0\leq i\leq d)$, set $\kappa_i = p^0_{ii}$
and note that $\kappa_i \neq 0$, since $R_i$ is non-empty. We
refer to $\kappa_i$ as the $i$-th valency of $Y$.

Let $Y = (X, \{R_i\}_{0\leq i\leq d})$ be a commutative symmetric
association scheme of class $d$, then the matrices $A_0,A_1,
...,A_d$ defined by
\begin{equation}\label{adj.}
    \bigl(A_{i})_{\alpha, \beta}\;=\left\{\begin{array}{c}
      \hspace{-2.3cm}1 \quad \mathrm{if} \;(\alpha,
    \beta)\in R_i, \\
      0 \quad \mathrm{otherwise} \quad \quad \quad(\alpha, \beta
    \in V) \\
    \end{array}\right.
\end{equation}
are adjacency matrices of $Y$ such that
\begin{equation}\label{ss}
A_iA_j=\sum_{k=0}^{d}p_{ij}^kA_{k}.
\end{equation}
From (\ref{ss}), it is seen that the adjacency matrices $A_0, A_1,
..., A_d$ form a basis for a commutative algebra \textsf{A} known
as the Bose-Mesner algebra of $Y$. This algebra has a second basis
$E_0,..., E_d$ (known as primitive idempotents of $Y$) so that
\begin{equation}\label{idem}
E_0 = \frac{1}{n}J, \;\;\;\;\;\;\ E_iE_j=\delta_{ij}E_i,
\;\;\;\;\;\;\ \sum_{i=0}^d E_i=I.
\end{equation}
where, $J$ is the all-one matrix in $\textsf{A}$. Let $P$ and $Q$
be the matrices relating the two bases for $\textsf{A}$:
$$
A_i=\sum_{i=0}^d P_{ij}E_j, \;\;\;\;\ 0\leq j\leq d,
$$
\begin{equation}\label{m2}
E_i=\frac{1}{n}\sum_{i=0}^d Q_{ij}A_j, \;\;\;\;\ 0\leq j\leq d.
\end{equation}
Then clearly
\begin{equation}\label{pq}
PQ=QP=nI.
\end{equation}
It also follows that
\begin{equation}\label{eign}
A_iE_j=P_{ij}E_j,
\end{equation}
which shows that the $P_{ij}$ (resp. $Q_{ij}$) is the $j$-th
eigenvalue (resp. the $j$-th dual eigenvalue ) of $A_i$ (resp.
$E_i$) and that the columns of $E_j$ are the corresponding
eigenvectors. Thus, $m_i=$rank$(E_i)$ is the multiplicity of the
eigenvalue $P_{ij}$ of $A_i$ (provided that $P_{ij}\neq P_{kj}$
for $k \neq i$). We see that $m_0=1, \sum_i m_i=n$, and
$m_i=$trace$E_i=n(E_i)_{jj}$ (indeed, $E_i$ has only eigenvalues
$0$ and $1$, so rank($E_k$) equals the sum of the eigenvalues).

Clearly, each non-diagonal (symmetric) relation $R_i$ of an
association scheme $Y=(V,\{R_i\}_{0\leq i\leq d})$ can be thought
of as the network $(V,R_i)$ on $V$, where we will call it the
underlying network of association scheme $Y$. In other words, the
underlying network $\Gamma=(V,R_1)$ of an association scheme is an
undirected connected network, where the set $V$ and $R_1$ consist
of its vertices and edges, respectively. Obviously replacing $R_1$
with one of the other relations such as $R_i$, for  $i\neq 0,1$
will also give us an underlying network $\Gamma=(V,R_i)$ (not
necessarily a connected network) with the same set of vertices but
a new set of edges $R_i$.

\subsection{Stratification} For an underlying network $\Gamma$, let
$W={\mathcal{C}}^n$ (with $n=|V|$) be the vector space over
$\mathcal{C}$ consisting of column vectors whose coordinates are
indexed by vertex set $V$ of $\Gamma$, and whose entries are in
$\mathcal{C}$. For all $\beta\in V$, let $\ket{\beta}$ denotes the
element of $W$ with a $1$ in the $\beta$ coordinate and $0$ in all
other coordinates. We observe $\{\ket{\beta} | \beta\in V\}$ is an
orthonormal basis for $W$, but in this basis, $W$ is reducible and
can be reduced to irreducible subspaces $W_i$, $i=0,1,...,d$,
i.e.,
\begin{equation}
W=W_0\oplus W_1\oplus...\oplus W_d,
\end{equation}
where, $d$ is diameter of the corresponding association scheme. If
we define
 $\Gamma_i(o)=\{\beta\in V:
(o, \beta)\in R_i\}$ for an arbitrary chosen vertex $o\in V$
(called reference vertex), then, the vertex set $V$ can be written
as disjoint union of $\Gamma_i(\alpha)$, i.e.,
 \begin{equation}\label{asso1}
 V=\bigcup_{i=0}^{d}\Gamma_{i}(\alpha).
 \end{equation}
In fact, the relation (\ref{asso1}) stratifies the network into a
disjoint union of strata (associate classes) $\Gamma_{i}(o)$. With
each stratum $\Gamma_{i}(o)$ one can associate a unit vector
$\ket{\phi_{i}}$ in $W$ (called unit vector of $i$-th stratum)
defined by
\begin{equation}\label{unitv}
\ket{\phi_{i}}=\frac{1}{\sqrt{\kappa_{i}}}\sum_{\alpha\in
\Gamma_{i}(o)}\ket{\alpha},
\end{equation}
where, $\ket{\alpha}$ denotes the eigenket of $\alpha$-th vertex
at the associate class $\Gamma_{i}(o)$ and
$\kappa_i=|\Gamma_{i}(o)|$ is called the $i$-th valency of the
network ($\kappa_i:=p^0_{ii}=|\{\gamma:(o,\gamma)\in
R_i\}|=|\Gamma_{i}(o)|$). For $0\leq i\leq d$, the unit vectors
$\ket{\phi_{i}}$ of Eq.(\ref{unitv}) form a basis for irreducible
submodule of $W$ with maximal dimension denoted by $W_0$. Since
$\{\ket{\phi_{i}}\}_{i=0}^d$ becomes a complete orthonormal basis
of $W_0$, we often write
\begin{equation}
W_0=\sum_{i=0}^d\oplus \textbf{C}\ket{\phi_{i}}.
\end{equation}

Let $A_i$ be the adjacency matrix of the underlying network
$\Gamma$. From the action of $A_i$ on reference state
$\ket{\phi_0}$ ($\ket{\phi_0}=\ket{o}$, with $o\in V$ as reference
vertex), we have
\begin{equation}\label{Foc1}
A_i\ket{\phi_0}=\sum_{\beta\in \Gamma_{i}(o)}\ket{\beta}.
\end{equation}
 Then by using (\ref{unitv}) and (\ref{Foc1}),
 we obtain
\begin{equation}\label{Foc2}
A_i\ket{\phi_0}=\sqrt{\kappa_i}\ket{\phi_i}.
\end{equation}

\section{Entanglement entropy between two parts of a network}
In order to calculate the entanglement entropy between two
equal parts in the graph (half first strata are in one subset and the other strata are in the second subset), we
introduce the following process:

First we want to generalize the Schur complement method. Suppose
we have the following matrix which is composed of block matrices.
\begin{equation}
V=\left(\begin{array}{ccc}
          V_{11}& V_{12} & 0\\
            V_{21}& V_{22} & V_{23}\\
            0 & V_{32} & V_{33}\\
          \end{array}\right)
\end{equation}
Then we do the generalized Schur complement transformation. This transformation is:
\begin{equation}
\left(\begin{array}{ccc}
          V_{11}& V_{12} & 0\\
            V_{21}& V_{22} & V_{23}\\
            0 & V_{32} & V_{33}\\
          \end{array}\right)=\left(\begin{array}{ccc}
          1& 0 & 0\\
            0& 1& V_{23}V_{33}^{-1}\\
            0 & 0 & 1\\
          \end{array}\right)\left(\begin{array}{ccc}
          V_{11}& V_{12} & 0\\
            V_{12}^T& V_{22}-V_{23}V_{33}^{-1}V_{32} & 0\\
            0 & 0 & V_{33}\\
          \end{array}\right)\left(\begin{array}{ccc}
          1& 0 & 0\\
            0& 1& 0\\
            0 & V_{33}^{-1}V_{32} & 1\\
          \end{array}\right)
\end{equation}
In our work, we will apply the generalized Schur complement method to the potential matrix in the stratification basis several times. So in the transformed matrix all of the blocks are scalar. Therefore the
potential matrix is transformed to a $2\times 2$ matrix finally.
\begin{equation}
V=\left(\begin{array}{cc}
         a_{11}&a_{12} \\
          a^T_{12}& a_{22}\\
          \end{array}\right)
\end{equation}
The wave function in this stage is
\begin{equation}
\psi(x,y)=A_g exp(-\frac{1}{2}(x\quad \quad
y)\left(\begin{array}{cc}
          a_{11}&a_{12}\\
            a_{12}& a_{22}\\
          \end{array}\right)\left(\begin{array}{c}
          x\\
            y\\
          \end{array}\right))
\end{equation}
by rescaling the variables $x$ and $y$:
$$\widetilde{x}=a_{11}^{1/2}x$$
$$\widetilde{y}=a_{22}^{1/2}y$$
the ground state wave function is transformed to
\begin{equation}
\psi(\widetilde{x},\widetilde{y})=A_g
exp(-\frac{1}{2}(\widetilde{x}\quad \quad
\widetilde{y})\left(\begin{array}{cc}
          1& \gamma\\
            \gamma & 1\\
          \end{array}\right)\left(\begin{array}{c}
          \widetilde{x}\\
            \widetilde{y}\\
          \end{array}\right))
\end{equation}
where
\begin{equation}\label{gamma}
\gamma=a_{11}^{-1/2}a_{12}a_{22}^{-1/2}
\end{equation}
So the ground state
wave function is
\begin{equation}\label{wave1}
\psi(\widetilde{x},\widetilde{y})=A_g
e^{-\frac{\widetilde{x}^2}{2}-\frac{\widetilde{y}^2}{2}-\gamma \widetilde{x}\widetilde{y}}
\end{equation}

Then we can use following identity to
calculate the schmidt number of this wave function,
\begin{equation}\label{identity}
\frac{1}{\pi^{1/2}}exp(-\frac{1+t^2}{2(1-t^2)}((x)^2+(y)^2))+\frac{2t}{1-t^2}x
y)=(1-t^2)^{1/2}\sum_n t^n \psi_n(x)\psi_n(y)
\end{equation}
In order to calculating the entropy, we apply a change of
variable as
$$1-t^2=\frac{2}{\nu+1}$$
$$t^2=\frac{\nu-1}{\nu+1}$$
So the above identity becomes
\begin{equation}\label{identity2}
\frac{1}{\pi^{1/2}}exp(-\frac{\nu}{2}((x)^2+(y)^2))+(\nu^2-1)^{1/2}x
y)=(\frac{2}{\nu+1})^{1/2}\sum_n (\frac{\nu-1}{\nu+1})^{n/2}
\psi_n(x)\psi_n(y)
\end{equation}
and the reduced density matrix is
\begin{equation}
\rho=\frac{2}{\nu+1}\sum_{n}(\frac{\nu-1}{\nu+1})^n |n\rangle
\langle n|
\end{equation}
the entropy is
\begin{equation}
S(\rho)=-\sum_n p_n log(p_n)
\end{equation}
where $p_n=\frac{2}{\nu+1}(\frac{\nu-1}{\nu+1})^n$
\begin{equation}
\sum_n p_n log(p_n)=log(\frac{2}{\nu+1})+\langle n \rangle
log(\frac{\nu-1}{\nu+1})
\end{equation}
and $\langle n \rangle = \frac{\nu -1}{2}$

\begin{equation}\label{Ent1}
S(\rho)=\frac{\nu +1}{2} log(\frac{\nu +1}{2})-\frac{\nu
-1}{2}log(\frac{\nu -1}{2})
\end{equation}
By comparing the wave function Eq.(\ref{wave1}) and the identity Eq.(\ref{identity2}) and
define the scale $\mu^2$, we conclude that
$$\nu=1 \times \mu^2$$
$$(\nu^2-1)^{1/2}=-\gamma \times \mu^2$$
After some straightforward calculation we obtain
\begin{equation}\label{gammanu}
\nu=(\frac{1}{1-\gamma^2})^{1/2}
\end{equation}
By above discussion we conclude that
$$e^{-\frac{(x)^2}{2}-\frac{(y)^2}{2}-\gamma xy}=\sum_n \lambda_{n}\psi_n(x)\psi_n(y)$$
where
$\lambda_{n}=(\frac{2}{\nu+1})^{1/2}(\frac{\nu-1}{\nu+1})^{n/2}$.

\subsection{Hypercube network $H(d,2)$}
The hypercube of dimension $d$ (known also as binary Hamming
scheme $H(d,2)$) is a network with $2^d$ nodes, each of which can
be labeled by a $d$-bit binary string. Suppose $x$ and $y$ are
words of $d$ bits each. The Hamming distance $d(x,y)$ between $x$
and $y$ is defined to be the number of places at which $x$ and $y$
differ. Two nodes on the hypercube described by bit strings $x$
and $y$ are connected by an edge if $|x-y|=1$, where $|x|$ is the
Hamming weight of $x$. The Hamming weight of a word $x$ is
defined to be the distance from the string of all zeros $d(x,0)$,
that is, the number of places at which $x$ is nonzero. In other
words, if $x$ and $y$ differ by only a single bit flip, then the
two corresponding nodes on the network are connected. Thus, each
of the $2^d$ nodes on the hypercube has degree $d$. For the
hypercube network with dimension $d$ we have $d+1$ strata with
\begin{equation}
\kappa_i=\frac{d!}{i!(d-i)!}\quad\quad 0\leq i\leq d-1
\end{equation}
Furthermore, the adjacency matrices of this network are given by
\begin{equation}
A_i =\sum_{perm} \underbrace{\sigma_x\otimes \sigma_x\ldots
\otimes \sigma_x}_i \otimes \underbrace{I_2\otimes \ldots \otimes
I_2}_{d-i} \quad\quad\quad i= 0,1,\ldots,d
\end{equation}
where, the summation is taken over all possible nontrivial
permutations.
The first stratification basis for this graph is
$$|\phi_0\rangle=|00\ldots0\rangle$$
$$|\phi_1\rangle=\frac{1}{\sqrt{C^d_1}}\sum_{i=1}^n|0\ldots0\underbrace{1}_i0\ldots0\rangle$$
$$\vdots$$
$$|\phi_k\rangle=\frac{1}{\sqrt{C^d_k}}\sum_{i_1\neq i_2\neq\ldots\neq i_k;i_1,\ldots,i_k=1}^d|0\ldots0\underbrace{1}_{i_1}0\ldots0\underbrace{1}_{i_2}0\ldots0\underbrace{1}_{i_k}0\ldots0\rangle$$
$$\vdots$$
\begin{equation}
|\phi_d\rangle=|11\ldots1\rangle
\end{equation}

The stratification basis of this graph can be written in the
basis of angular momentum by considering $J=d/2$, So
\begin{equation}
|Jm\rangle=\frac{1}{\sqrt{\frac{n!}{n_0!n_1!}}}\sum|\underbrace{00\ldots0}_{n_0}\underbrace{11\ldots1}_{n_1}\rangle
\end{equation}
where $n_0+n_1=1$ and $m=\frac{n_0-n_1}{2}$. So for example the
first stratum ($|00\ldots0\rangle$) is equivalent to the
$|JJ\rangle$.

So the adjacency matrix in the first stratification basis, is the
matrix representation of the first component of angular momentum
$J=\frac{d}{2}$ as
\begin{equation}\label{AJx}
A=\left(\begin{array}{cccccc}
         0& \sqrt{d} & 0&0&\ldots &0\\
            \sqrt{d}& 0 & \sqrt{2(d-1)}&0&\ldots &0\\
            0 & \sqrt{2(d-1)} & 0& \sqrt{3(d-2)}&\ldots &0\\
            \vdots & \ddots & \ddots& \ddots & \vdots & \vdots \\
            0&0&\ldots & \sqrt{2(d-1)} &0 &\sqrt{d}\\
            0 & 0 & 0 & \ldots & \sqrt{d} & 0\\
          \end{array}\right)
\end{equation}

The above equation gives the first block of adjacency matrix in the stratification basis.

The other stratification basis which give the other blocks of adjacency matrix, may be obtained by applying the lowering
operator to the $|JJ\rangle\equiv |00\ldots0\rangle$, So it will
be in the form
$$\sum_i a_i |0\ldots0\underbrace{1}_i0\ldots0\rangle$$
The action of raising operator on the above state will be zero,So
we have $\sum a_i=0$ Therefore the number of degeneracy in second
stratum is $C^d_1-C^d_0\equiv d-1$ The highest state in the
second stratum is
\begin{equation}\label{Jm2}
|J=d/2-1,m=d/2-1\rangle_{k=1,\ldots,d-1}=\frac{1}{\sqrt{d}}\sum_{l=0}^d
w^{kl} |0\ldots0\underbrace{1}_{l}0\ldots0\rangle
\end{equation}
The second block of adjacency matrix, is similarly the
representation of first component of angular momentum matrix, but
the dimension has decreased to $d'=d-2$ (i.e. $J=\frac{d-2}{2}$)
and degeneracy is $d-1$. From the fourier transformation
($w^d=1$), the sum of the coefficients in the above equation is
zero, because
$$1+w^k+(w^k)^2+\ldots +(w^k)^{d-1}=0$$
Now we want to construct the third block of adjacency matrix. If
we apply the lowering operator on the Eq.(\ref{Jm2}), the result
will be
\begin{equation}
\sum_{i_1\neq
i_2}a_{i_1i_2}|0\ldots0\underbrace{1}_{i_1}0\ldots0\underbrace{1}_{i_2}0\ldots0\rangle
\end{equation}
The number of coefficients $a_{i_1i_2}$ in the above is $C^d_2$.
The condition $\sum_{i_1\neq i_2}a_{i_1i_2}=0$, produces $C^d_1=d$
equations, So the total degeneracy of the third stratum is
$C^d_2-C^d_1$.\\Table 1 show the angular momentum corresponding to the blocks of adjacency matrix and their degeneracies.

\begin{table}
\begin{tabular}{|c|c|}
\hline
Adjacency matrix & degeneracy \\
\hline $J=\frac{d}{2}$   & $C_0^d$  \\
\hline $J=\frac{d-2}{2}$   & $C_1^d-C_0^d$ \\
\hline $J=\frac{d-4}{2}$   & $C_2^d-C_1^d$ \\
\hline $J=\frac{d-6}{2}$   & $C_3^d-C_2^d$ \\
\hline $\vdots$   & $\vdots$ \\
\hline $J=\sigma_x \quad\quad d\quad odd$   & $C^d_{\frac{d+1}{2}-1}-C^d_{\frac{d+1}{2}-2}$ \\
\hline singlet $0 \quad\quad d \quad even$   & $C^d_{\frac{d}{2}}-C^d_{\frac{d}{2}-1}$ \\
\hline
\end{tabular}
\caption{\label{angular momentum}The angular momentum corresponding to the blocks of adjacency matrix and their degeneracies. }
\end{table}

For example the Hypercube network $H(3,2)$ in FIG.1, has four strata. The vertex labeled $1$ is in the first stratum, the vertices labeled $2,3,4$ are in the second stratum, the vertices labeled $5,6,7$ are in the third stratum and the vertex labeled $8$ is in the fourth stratum. The first block of adjacency matrix in the stratification basis for this graph is obtained from Eq.(\ref{AJx}) as
\begin{equation}
\left(\begin{array}{cccc}
         0& \sqrt{3} & 0&0\\
            \sqrt{3}& 0 & 2&0\\
            0 & 2 & 0& \sqrt{3}\\
            0 & 0 & \sqrt{3} & 0\\
          \end{array}\right)
\end{equation}
The other blocks of adjacency matrix of $H(3,2)$ are obtained from Table 1 as
\begin{equation}
A=\left(\begin{array}{cc}
        0&1\\
         1&0\\
          \end{array}\right)
\end{equation}
and its degeneracy is $C_1^3-C_0^3=2$. So the total adjacency matrix of $H(3,2)$ in the stratification basis is
\begin{equation}
\left(\begin{array}{cccccccc}
         0& \sqrt{3} & 0&0 & 0 & 0 & 0 & 0\\
            \sqrt{3}& 0 & 2&0& 0 & 0 & 0 & 0\\
            0 & 2 & 0& \sqrt{3}& 0 & 0 & 0 & 0\\
            0 & 0 & \sqrt{3} & 0& 0 & 0 & 0 & 0\\
            0 & 0 & 0 & 0 & 0 & 1 & 0 & 0 \\
            0 & 0 & 0 & 0 & 1 & 0 & 0 & 0 \\
            0 & 0 & 0 & 0 & 0 & 0 & 0 & 1 \\
            0 & 0 & 0 & 0 & 0 & 0 & 1 & 0 \\
          \end{array}\right)
\end{equation}

\subsubsection{Entanglement entropy in Hypercube network between two equal parts: First half and second half strata}
Now we want to calculate bipartite entanglement between two equal parts of Hypercube network. First we consider the $H(d,2)$ network with odd $d$. The potential matrix is
\begin{equation}
V=(1+2gd)I_{(d+1)\times(d+1)}-2gA_{(d+1)\times(d+1)}
\end{equation}
where $A_{(d+1)\times(d+1)}$ is the adjacency matrix in the form
of Eq.(\ref{AJx}).
\begin{equation}
V=\left(\begin{array}{cccccc}
         1+2dg& -2g\sqrt{d} & 0&0&\ldots &0\\
            -2g\sqrt{d}& 1+2dg & -2g\sqrt{2(d-1)}&0&\ldots &0\\
            0 & -2g\sqrt{2(d-1)} & 1+2dg& -2g\sqrt{3(d-2)}&\ldots &0\\
            \vdots & \ddots & \ddots& \ddots & \vdots & \vdots \\
            0&0&\ldots & -2g\sqrt{2(d-1)} &1+2dg &-2g\sqrt{d}\\
            0 & 0 & 0 & \ldots & -2g\sqrt{d} & 1+2dg\\
          \end{array}\right)
\end{equation}
Then by using Schur complement method
to the first two blocks and two last blocks, we have
\begin{equation}
V=\left(\begin{array}{cccccc}
         1+2dg& 0 & 0&0&\ldots &0\\
            0& 1+2dg-\frac{4g^2d}{1+2dg} & -2g\sqrt{2(d-1)} & 0 &\ldots &0\\
            0 & -2g\sqrt{2(d-1)} & 1+2dg& -2g\sqrt{3(d-2)}&\ldots &0\\
            \vdots & \ddots & \ddots& \ddots & \vdots & \vdots \\
            0&0&\ldots & -2g\sqrt{2(d-1)} &1+2dg-\frac{4g^2d}{1+2dg} &0\\
            0 & 0 & 0 & \ldots & 0 & 1+2dg\\
          \end{array}\right)
\end{equation}
After applying the generalized Schur complement method to the potential matrix alternatively, we have one $2\times2$ matrix finally. The elements of this $2\times2$ matrix is
$$a_{12}=-2g\sqrt{(\frac{d+1}{2})(\frac{d+1}{2})}=-g(d+1)$$
\begin{equation}
a_{11}=a_{22}=1+2dg-\frac{4g^2(\frac{d-1}{2})(\frac{d-3}{2})}{1+2dg-\frac{4g^2(\frac{d-3}{2})(\frac{d-5}{2})}{\frac{\vdots}{1+2dg-\frac{4g^2(2(d-1))}{1+2dg-\frac{4g^2d}{1+2dg}}}}}
\end{equation}
So from the Eq.(\ref{gamma}) the parameter $\gamma_1$ ($\gamma_i$
for $i$th stratum) will be
\begin{equation}
\gamma_1=\frac{\frac{(d+1)}{2}}{d+\frac{1}{2g}-\frac{(\frac{d-1}{2})(\frac{d-3}{2})}{d+\frac{1}{2g}-\frac{(\frac{d-3}{2})(\frac{d-5}{2})}{\frac{\vdots}{d+\frac{1}{2g}-\frac{(2(d-1))}{d+\frac{1}{2g}-\frac{d}{d+\frac{1}{2g}}}}}}}
\end{equation}
So the parameter $\gamma_1$ can be written as
\begin{equation}
\gamma_1=\frac{d+1}{2}\frac{Q_{n-1}(x)}{Q_n(x)}
\end{equation}
The parameter $x$ defined as $x=d+\frac{1}{2g}$.
The recursive polynomials $Q_{n}(x)$ are defined as

$$Q_{1}(x)=x\quad,\quad Q_{2}(x)=x^2-\omega_1x$$
\begin{equation}
Q_{n}(x)=xQ_{n-1}(x)-\omega_{n-1}Q_{n-2}(x),\quad\quad\quad \omega_i=i(d-i+1),\quad\quad i=1,2,\ldots,\frac{d-1}{2}
\end{equation}
for $n\geq3$. We should apply this method to all potential
matrices of all strata. So parameters $\nu_i$ are calculated from
all of parameters $\gamma_i$ from Eq.(\ref{gammanu}). Finally the entropy of
entanglement of each strata, i.e. $S(\rho_i)$, is obtained from
Eq.(\ref{Ent1}). So the total entropy is
\begin{equation}
S(\rho)=\sum_i S(\rho_i)
\end{equation}
\subsubsection{Entanglement entropy in Hypercube network between two equal parts: The connection matrix between two parts is Identity}
Suppose that we divide the $H(d,2)$ graph into two equal parts,
such that each of parts is another Hypercube graph with smaller
dimension $H(d-1,2)$, and the connection matrix between two parts,
is Identity matrix. So the adjacency matrix is
\begin{equation}
A_{H(d,2)}=\left(\begin{array}{cc}
         A_{H(d-1,2)}&I_{2^{d-1}\times 2^{d-1}} \\
          I_{2^{d-1}\times 2^{d-1}} & A_{H(d-1,2)}\\
          \end{array}\right)
\end{equation}
Then the potential matrix is
\begin{equation}
V_{H(d,2)}=\left(\begin{array}{cc}
         (1+2dg)I-2gA & -2gI \\
          -2gI & (1+2dg)I-2gA \\
          \end{array}\right)
\end{equation}
First we diagonalize the Matrix $(1+2dg)I-2gA$, by a local transformation.
This transformation dosen't change the Identity matrix, so the result will be
\begin{equation}
V_{H(d,2)}=\left(\begin{array}{cc}
         (1+2dg)I-2gD &-2gI \\
          -2gI & (1+2dg)I-2gD \\
          \end{array}\right)
\end{equation}
where
\begin{equation}
D=\left(\begin{array}{cccc}
         \lambda_1 & 0 & \ldots & 0 \\
         0 & \lambda_2 & \ldots & 0 \\
         \vdots & \vdots & \ddots & \vdots \\
         0 & 0 & \ldots & \lambda_{2^{d-1}} \\
          \end{array}\right)
\end{equation}
and $\lambda_i$s are the eigenvalues of adjacency matrix of
Hypercube network $H(d-1,2)$. The eigenvalues of adjacency
matrices $A_i$s are given from the Krawtchouk polynomials as
\begin{equation}
K_l(x)=\sum_{i=0}^{l} \left(\begin{array}{c}
         x\\
          i\\
          \end{array}\right)\left(\begin{array}{c}
        d-x\\
          l-i\\
          \end{array}\right)(-1)^i
\end{equation}
So the eigenvalues of adjacency matrix $A_1$ of graph Hypercube $H(d-1,2)$ are given from
\begin{equation}
K_1(i)=(d-1)-2i
\end{equation}
And the corresponding degeneracies are given from
\begin{equation}
n_i=\frac{(d-1)!}{2^{d-1}i!(d-1-i)!}
\end{equation}
Then the parameter $\gamma_i$ for Schmidt number is
\begin{equation}
\gamma_i=\frac{2g}{1+2g(d'+1-\lambda_i)}
\end{equation}
Where the quantity $d'-\lambda_i$ is the Laplacian matrix of Hypercube $H(d',2)=H(d-1,2)$
This kind of partition in Hypercube networks, is equivalent to the minimum entanglement entropy.
\subsubsection{Entanglement entropy in Hypercube network between two equal parts: even strata are in first subset and odd strata are in second subset}
The entanglement entropy and Schmidt numbers are considered between
two equal parts of hypercube network, that even strata are in
first subset and odd strata are in second subset. In this kind of
partitioning we should consider two cases as $d=odd$
and $d=even$.
\\\textbf{Case I($d$ is odd):}In this case the potential matrix can be written
as following form:
\begin{equation}
V_{11}=V_{22}=I_{\frac{d+1}{2}\times\frac{d+1}{2}}(1+2gd)
\end{equation}
and the connection matrix of potential matrix is:
\begin{equation}
V_{12}=\left(%
\begin{array}{ccccc}
  -2g\sqrt{d} & 0 & 0 & ... & 0 \\
  -2g\sqrt{2(d-1)} & -2g\sqrt{3(d-2)} & 0 & ... & 0 \\
  0 & -2g\sqrt{3(d-2)} & -2g\sqrt{4(d-3)} & ... & 0 \\
  \vdots & \vdots & \vdots & \ddots & \vdots \\
  0 & ... & 0 & -2g\sqrt{2(d-1)} & -2g\sqrt{d} \\
\end{array}%
\right)_{\frac{d+1}{2}\times\frac{d+1}{2}}
\end{equation}
Therefore,we have $\frac{d+1}{2}$ parameters $\gamma$
$$\gamma_{1}=\frac{2g}{1+2gd}$$
$$\gamma_{2}=\frac{(2+4)g}{1+2gd}$$
$$\gamma_{3}=\frac{(2+4\times2)g}{1+2gd}$$
$$\vdots$$
\begin{equation}
\gamma_{\frac{d+1}{2}}=\frac{(2+4\frac{d-1}{2})g}{1+2gd}
\end{equation}
\textbf{Case II($d$ is even):}In this case the potential matrix
can be written as following
\begin{equation}
V_{11}=I_{\frac{d}{2}+1\times\frac{d}{2}+1}(1+2gd)
\end{equation}
\begin{equation}
V_{22}=I_{\frac{d}{2}\times\frac{d}{2}}(1+2gd)
\end{equation}
and the connection matrix of potential matrix is
\begin{equation}
V_{12}=\left(%
\begin{array}{ccccc}
  -2g\sqrt{d} & 0 & ... & 0 \\
  -2g\sqrt{2(d-1)} & -2g\sqrt{3(d-2)} & ... & 0 \\
  0 & -2g\sqrt{3(d-2)}  & ... & 0 \\
  \vdots & \vdots  & \ddots & \vdots \\
  0 & ... & -2g\sqrt{2(d-1)} & -2g\sqrt{d} \\
\end{array}%
\right)_{\frac{d}{2}\times\frac{d}{2}+1}
\end{equation}
Therefore, we have $\frac{d}{2}$ parameters $\gamma$
$$\gamma_{1}=\frac{4g}{1+2gd}$$
$$\gamma_{2}=\frac{(4+4)g}{1+2gd}$$
$$\gamma_{3}=\frac{(4+4\times2)g}{1+2gd}$$
$$\vdots$$
\begin{equation}
\gamma_{\frac{d}{2}}=\frac{(4+4(\frac{d}{2}-1))g}{1+2gd}
\end{equation}
In the next section we will show that this kind of partitioning in Hypercube
networks, gives the maximum entanglement entropy.
\section{Numerical result for entanglement entropy}
\subsection{Entanglement entropy between all kinds of equal parts
in Hypercube H(3,2) network} In this section we separated
Hypercube $H(3,2)$ network into to equal parts, then calculated
the entanglement entropy between all kinds of bisection.

\setlength{\unitlength}{0.75cm}
\begin{picture}(6,5)
\linethickness{0.075mm}

\put(6,1){\circle*{0.2}} \put(6,3){\circle*{0.2}}
\put(8,1){\circle*{0.2}} \put(8,3){\circle*{0.2}}

\put(7,2){\circle*{0.2}} \put(7,4){\circle*{0.2}}
\put(9,2){\circle*{0.2}} \put(9,4){\circle*{0.2}}

\put(6,1){\line(1,0){2}} \put(6,1){\line(0,1){2}}
\put(6,3){\line(1,0){2}} \put(8,1){\line(0,1){2}}

\put(7,2){\line(1,0){2}} \put(7,2){\line(0,1){2}}
\put(7,4){\line(1,0){2}} \put(9,2){\line(0,1){2}}

\put(6,1){\line(1,1){1}} \put(6,3){\line(1,1){1}}
\put(8,3){\line(1,1){1}} \put(8,1){\line(1,1){1}} \put(5.6,1){$1$}
\put(5.6,3){$2$} \put(6.6,2){$3$} \put(8.3,1){$4$}
\put(9.2,2){$5$} \put(7.8,3.2){$6$} \put(6.6,4){$7$}
\put(9.4,4){$8$}

\put(0,-1){\footnotesize  FIG I: Hypercube $H(3,2)$ graph.}
\end{picture}
\newline
\newline
\newline
So there are $\frac{1}{2}\left(\begin{array}{c}
          8\\
            4\\
          \end{array}\right)=35$ kinds of partitioning graph into equal parts. These $35$ kinds of partitioning only have $6$ amount of entanglement entropy as shown in table II. The maximum entanglement entropy in this section is for subset: $(1, 5, 6, 7)$. It is for the case that we choose the vertices of first, and third strata.
The minimum entanglement entropy in this section are for subsets:
$(1, 2, 3, 7),(1, 2, 4, 6),(1, 3, 4, 5)$. These are the cases that
the connection matrix between two equal parts is identity. In
these cases there are the minimum number of edges between two equal
parts.
\begin{table}
\begin{tabular}{|c|c|}
\hline
set & partitions \\
\hline $1$   & $(1,5,6,7)$  \\
\hline       & $(1,2,3,8),(1,2,4,8),(1,2,5,6),(1,2,5,7)$ \\
       $2$   & $(1,3,4,8),(1,3,5,6),(1,3,6,7),(1,4,5,7)$ \\
             & $(1,4,6,7),(1,5,6,8),(1,5,7,8),(1,6,7,8)$ \\
\hline $3$   & $(1,2,5,8),(1,3,6,8),(1,4,7,8)$ \\
\hline       & $(1,2,3,5),(1,2,3,6),(1,2,4,5),(1,2,4,7)$ \\
       $4$   & $(1,2,6,8),(1,2,7,8),(1,3,4,6),(1,3,4,7)$ \\
             & $(1,4,5,8),(1,3,5,8),(1,3,7,8),(1,4,6,8)$ \\
\hline $5$   & $(1,2,3,4),(1,2,6,7),(1,3,5,7),(1,4,5,6)$ \\
\hline $6$   & $(1,2,3,7),(1,2,4,6),(1,3,4,5)$ \\
\hline
\end{tabular}
\caption{\label{entropy} All equal subsets in $H(3,2)$. The
entanglement entropy for these subsets are
$S_1>S_2>S_3>S_4>S_5>S_6$.}
\end{table}
\subsection{Entanglement entropy between all kinds of equal parts
in Hypercube H(4,2) network} In this section we separated
Hypercube $H(4,2)$ network into to equal parts, then calculated
the entanglement entropy between all kinds of bisection
numerically.
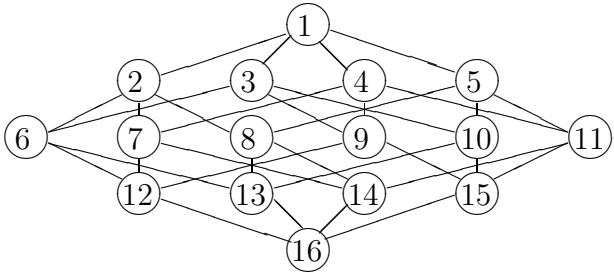
\newline
\newline
\setlength{\unitlength}{0.75cm}
\begin{picture}(6,4)
\linethickness{0.075mm}

\put(4.7,-0.2){$16$} \put(1.7,0.8){$12$} \put(3.7,0.8){$13$}
\put(5.7,0.8){$14$} \put(7.7,0.8){$15$}

\put(-0.2,1.8){$6$} \put(1.8,1.8){$7$} \put(3.8,1.8){$8$}
\put(5.8,1.8){$9$} \put(7.7,1.8){$10$} \put(9.7,1.8){$11$}

\put(1.8,2.8){$2$} \put(3.8,2.8){$3$} \put(5.8,2.8){$4$}
\put(7.8,2.8){$5$}

\put(4.8,3.8){$1$}

\put(5,0){\circle{0.8}}

\put(2,1){\circle{0.8}} \put(4,1){\circle{0.8}}
\put(6,1){\circle{0.8}} \put(8,1){\circle{0.8}}

\put(0,2){\circle{0.8}} \put(2,2){\circle{0.8}}
\put(4,2){\circle{0.8}} \put(6,2){\circle{0.8}}
\put(8,2){\circle{0.8}} \put(10,2){\circle{0.8}}

\put(2,3){\circle{0.8}} \put(4,3){\circle{0.8}}
\put(6,3){\circle{0.8}} \put(8,3){\circle{0.8}}

\put(5,4){\circle{0.8}}

\put(2,1.4){\line(0,1){0.2}} \put(2,2.4){\line(0,1){0.2}}
\put(8,1.4){\line(0,1){0.2}} \put(8,2.4){\line(0,1){0.2}}
\put(4,1.4){\line(0,1){0.2}} \put(6,2.4){\line(0,1){0.2}}

\put(2.4,3.1){\line(3,1){2.2}} \put(5.4,3.9){\line(3,-1){2.2}}

\put(4.2,3.3){\line(1,1){0.5}} \put(5.2,3.7){\line(1,-1){0.5}}

\put(2.4,0.9){\line(3,-1){2.3}} \put(4.4,0.9){\line(1,-1){0.5}}
\put(5.2,0.3){\line(1,1){0.5}} \put(5.3,0.2){\line(3,1){2.3}}

\put(0.4,2.1){\line(2,1){1.3}} \put(0.4,1.9){\line(2,-1){1.3}}
\put(0.4,2.1){\line(4,1){3.2}} \put(0.4,1.9){\line(4,-1){3.2}}

\put(9.6,2.1){\line(-2,1){1.3}} \put(9.6,1.9){\line(-2,-1){1.3}}
\put(9.6,2.1){\line(-4,1){3.2}} \put(9.6,1.9){\line(-4,-1){3.2}}

\put(2.4,2.1){\line(4,1){3.2}} \put(2.4,1.9){\line(4,-1){3.2}}

\put(4.4,2.1){\line(4,1){3.2}} \put(4.4,1.9){\line(2,-1){1.3}}
\put(3.6,2.1){\line(-2,1){1.3}}

\put(6.4,1.9){\line(2,-1){1.3}} \put(5.6,2.1){\line(-2,1){1.3}}
\put(5.6,1.9){\line(-4,-1){3.2}}

\put(7.6,2.1){\line(-4,1){3.2}} \put(7.6,1.9){\line(-4,-1){3.2}}
\end{picture}
\\In this case, there are $6435$ kinds of partitioning in a way that
two parts have equal vertices. Numerical calculations show that
some of these $6435$ sets have the same entropy, such that there
are $55$ kinds of different values for entanglement entropies. The
number of copies of these $55$ sets are
$1,4(2),6,12,16,24(2),32(4),48(8),72,96(12),192(20),384(2)$. The
maximum entanglement entropy in this section is for subset: $(1,
6, 7, 8, 9, 10, 11, 16)$. It is for the case that we choose the
vertices of first, third and last strata. In this case there are
the maximum edges between two equal parts. The minimum
entanglement entropy in this section are for subsets:
$$(1, 2, 3,4, 6, 7, 9, 12),(1, 2, 3, 5, 6, 8, 10, 13),(1, 2, 4, 5, 7, 8, 11, 14),(1, 3, 4, 5, 9, 10, 11, 15)$$
These are the cases that the connection matrix between two equal
parts is identity. In these cases there are the minimum number of
edges between two parts.
\section{Conclusion}
The entanglement entropy is obtained between two parts in the
hypercube networks which their nodes are considered as quantum
harmonic oscillators. The generalized Schur complement method is
used to calculate the Schmidt numbers and entanglement entropy
between two parts of hypercube graph. Analytically, entanglement
entropy in three special partitioning are calculated.
\\we investigated some kinds of partitioning in Hypercube network. So, the
ordering of entanglement entropies for these partitioning are
given. By using this result, we can conjecture which partitioning have the
minimum and maximum entanglement entropy.
\\One expects that the entanglement entropy and Schmidt numbers
can be calculated in Hamming networks.

\end{document}